\documentclass[9pt,twocolumn,twoside]{osajnl}

\journal{optica} 

\setboolean{shortarticle}{false}

\title{Engineered Raman Lasing in Photonic Integrated Chalcogenide Microresonators}

\author[1,5]{Yufei Huang}
\author[1,5]{Di Xia}
\author[1]{Pingyang Zeng}
\author[1]{Jiaxin Zhao}
\author[1]{Zelin Yang}
\author[4]{Suwan Sun}
\author[1]{Liyang Luo}
\author[1]{Guiying Hu}
\author[1]{Dong Liu}
\author[1]{Yufei Li}
\author[4]{Hairun Guo}
\author[1,2,*]{Bin Zhang}
\author[1,2,3,*]{Zhaohui Li}

\affil[1]{Guangdong Provincial Key Laboratory of Optoelectronic Information Processing Chips and Systems, School of Electrical and Information Technology, Sun Yat-sen University, Guangzhou 510275, China}
\affil[2]{Key Laboratory of Optoelectronic Materials and Technologies, Sun Yat-sen University, Guangzhou 510275, China}
\affil[3]{Southern Marine Science and Engineering Guangdong Laboratory (Zhuhai), Zhuhai, 519000, China}
\affil[4]{Key Laboratory of Specialty Fiber Optics and Optical Access Networks, Shanghai University, Shanghai 200444, China}
\affil[5]{these authors contributed equally to this work}

\affil[*]{Corresponding author: zhangbin5@mail.sysu.edu.cn, lzhh88@mail.sysu.edu.cn }

\begin{abstract}

Chalcogenide glass (ChG) is an attractive material for integrated nonlinear photonics due to its wide transparency and high nonlinearity, and its capability of being directly deposited and patterned on Silicon wafer substrates. It has a singular Raman effect among amorphous materials. Yet, the Raman lasing performance in high quality and chip integrated ChG microresonators remains unexplored. Here, we demonstrate an engineered Raman lasing dynamic based on home developed photonic integrated high-$Q$ ChG microresonators. With a quality factor above $10^{6}$, we achieve the record-low lasing threshold 3.25 mW among integrated planar photonic platforms. Both the single-mode Raman lasers and a broadband Raman-Kerr comb are observed and characterized, which is dependent on the dispersion of our flexible photonic platform and engineered via tuning the waveguide geometric size. The tunability of such a chip-scale Raman laser is also demonstrated through tuning the pump wavelength and tuning the operating temperature on the chip. This allows for the access of single-mode lasing at arbitrary wavelengths in the range 1615-1755 nm. Our results may contribute to the understanding of rich Raman and Kerr nonlinear interactions in dissipative and nonlinear microresonators, and on application aspect, may pave a way to chip-scale efficient Raman lasers that is highly desired in spectroscopic applications in the infrared.

\end{abstract}

\setboolean{displaycopyright}{true}

\begin{document}

\maketitle
\section{Introduction}

Stimulated Raman scattering (SRS) is well known as an effective means to extend the spectral coverage of conventional semiconductor and rare-earth-doped laser sources \cite{rong2008cascaded,takahashi2013micrometre,bernier20143,Zhange2101605118}, in which a long-wavelength signal (i.e. the Stokes radiation) can be generated from the initial pump wave, with a frequency offset equal to the molecular vibrational frequency underlying the material \cite{shen2020raman,PhysRevLett.125.183901}. Potential applications of Raman lasing include optical amplification, spectroscopic sensing, archaeology, and clinical diagnosis \cite{shen2020raman,rong2005all,bernier2013mid}. 

\begin{figure*}[ht!]
\centering\includegraphics[width=\linewidth]{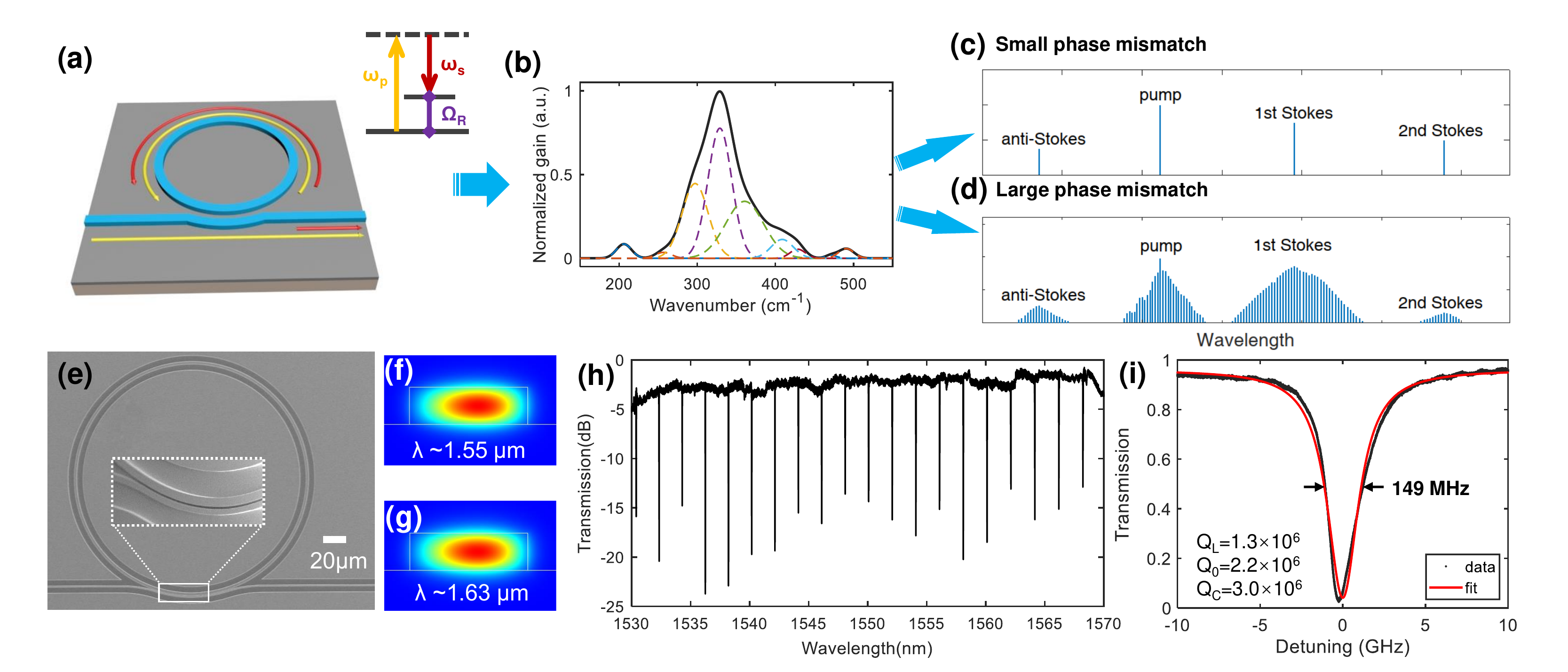}
\caption{
\textbf{Chip-scale chalcogenide microresonators.} (a) A schematic ChG microresonator and a sketch of the energy level system of the Raman effect. (b) Measured Raman spectrum of ChG film with a thickness of 0.8 $\mu$m. (c) Illustration of the single-mode and cascaded Raman laser in the microresonator with FWM phase matching (cf. Eq. \ref{eq:1}, \ref{eq:2}). (d) The Raman-Kerr comb state in the presence of FWM phase mismatch. (e) The SEM image of a GeSbS microresonator, together with the image on the coupling region. The cross-section of the GeSbS waveguide is 2.4 $\mu$m × 0.8 $\mu$m, and a radius of 100 $\mu$m. The microring has a 1.3 $\mu$m wide bus waveguide and a 35° pulley-style coupler with a critical coupling gap of ca. 550 nm. Simulated field distribution at (f) the pump wave and (g) the first Stokes wave, respectively. The microring can support good spatial overlap between the pump and Stokes mode. (h) Measured transmission spectra and (i) Resonance of fundamental TE modes of the microring without an Erbium-doped fiber amplifier (EDFA).}
\label{fig:1}
\end{figure*}

Recent development has been on photonic integrated platforms, where Raman lasers with high compactness and low energy consumption are enabled \cite{zhou2015chip,ahmadi2021widely,latawiec2015chip,liu2017integrated}. Indeed, photonic integrated Raman lasers have been experimentally demonstrated in silicon racetrack resonators \cite{rong2008cascaded,zhou2015chip,ahmadi2021widely}, silicon photonic crystals \cite{takahashi2013micrometre}, on-chip diamond resonators \cite{latawiec2015chip}, photonic aluminum nitride (AlN) microresonators \cite{liu2017integrated}, lithium niobate microresonators (LN) \cite{yu2020raman} and silica microcavities \cite{kippenberg2004ultralow}. Yet, these platforms intrinsically have shortcomings that limit the performance of the Raman lasing. For instance, silicon has severe two-photon absorption (TPA) and free-carrier absorption (FCA), and silica suffers from a much weaker Raman gain coefficient than silicon \cite{rong2008cascaded,takahashi2013micrometre,rong2005all,zhou2015chip}. On fabrication aspect, diamond, LN and crystalline AlN need to be co-integrated on a Silicon substrate either by the “wafer bonding” or carefully controlled crystal growth processes, which is non-trivial to implement \cite{latawiec2015chip,liu2017integrated,yu2020raman}. Moreover, the Raman gain in crystalline materials is always narrow-band, which requires precise control on the crystal orientation as well as on the cut. Control on the cavity resonant frequency is also required to align with the Raman mode. 

Among amorphous materials, chalcogenide glass (ChG) has a singular Raman and Kerr nonlinearity while being resistant to TPA and FCA \cite{eggleton2011chalcogenide,lin2017chalcogenide,eggleton2019brillouin,zhang2021chip,lpor.202000481,Zhu:19}. It has a wide transparency from the visible to the footprint infrared region (up to ca. 25 $\mu$m) and has a tailorable nonlinearity by adjusting the alloy composition \cite{yu2014broadband,petersen2014mid,PhysRevMaterials.3.035601}. Such properties have led to ultra-broadband and efficient supercontinuum generation in ChG based optical fibers and waveguides \cite{yu2014broadband,petersen2014mid,zhang2016high}, and strong SRS within a wide wavelength range up to the mid-infrared (above 4 $\mu$m) has been demonstrated \cite{bernier20143,bernier2013mid,Tuniz:08,Peng:19}. Of particular interest is photonic chip integrated ChG devices \cite{Giacoumidis:18}, which could largely boost the performance of SRS and the Raman lasing. Indeed, ChG can adhere to the Silicon wafer substrate by thermal evaporation at a low glass transition temperature (\textit{$T_g$}, less than $300^{\circ}\mathrm{C}$), free from the bonding or surface modification processes \cite{lin2017chalcogenide,li2014integrated,serna2018nonlinear,Jean:20,kim2020universal}. Being amorphous, the material Raman gain becomes broadband, which enables a flexible design on the free spectral range (FSR) of integrated microresonators \cite{latawiec2015chip,liu2017integrated}. To date, single-mode Raman lasing has been implemented in ChG microspheres \cite{andrianov2021tunable,vanier2014cascaded}. Yet, there remain challenges in the implementation of high-quality and chip integrated ChG microresonators. Usually, ChG films prepared with Arsenic (As)-based components are prone to oxidation and photorefraction. The on-chip ChG-based devices also suffer from laser-induced damage when being exposed to a pump power of tens mW \cite{broaddus2009silicon,choi2013photo}. Regarding the quality of ChG microresonators, the reported $Q$-factors to date remain insufficient to support high-efficient nonlinear interactions as well as the SRS \cite{du2018chip}. 

In this work, we overcome the above challenges by developing a photonic integrated high-$Q$ ChG microresonator platform and demonstrate a highly efficient chip-scale Raman lasing at the telecommunication $L$ band. The integrated microresonators are based on environment-friendly Ge$_{25}$Sb$_{10}$S$_{65}$ (GeSbS), which has strong Kerr nonlinearity, high Raman gain, high damage threshold and superior thermal stability \cite{shang2021chip,9431687,jace.14025}. The intrinsic $Q$-factor of our microresonators is as high as $2.2\times10^{6}$, and the threshold power for the Raman lasing is as low as 3.25 mW. We demonstrate both the single-mode and the cascaded Raman lasing, together with a broadband Raman-Kerr comb \cite{suzuki2018broadband}, by engineering the dispersion and the four-wave-mixing (FWM) phase matching condition underlying the microresonator. The tunability of the Raman laser is also demonstrated to cover a wavelength range over 140 nm. Our work marks, to the best of our knowledge, the first observation of engineered and widely tunable Raman lasers in photonic chip integrated platforms, which may contribute not only to the fundamental physics of Raman lasing in dissipative Kerr resonators, but also to spectroscopic applications by e.g. supporting a hyperspectral imaging.

\section{Device fabrication and characterization}
\label{sec:design}

The photonic integrated ChG waveguides and microresonators are fabricated based on an improved subtractive nano-fabrication process that allows for stable and high-quality ChG microresonators \cite{zhang2021chip,9431687}. Initially, GeSbS glass is synthesized from high purity elements, including 6N Germanium (Ge), 6N Antimony (Sb) and 6N Sulfur (S), by the conventional melt-quenching technique \cite{zhang2016high,jace.14025}, which is further purified by our home developed physical and chemical purification technique \cite{zhang2016high}. The high purity GeSbS is then deposited on a Silicon wafer with a 3-$\mu$m SiO$_2$ layer via thermal evaporation, forming a film of 0.8-$\mu$m thickness. Next, the film is patterned by electron-beam lithography. Afterwards, the pattern is etched in an inductively coupled plasma reactive ion etcher (ICP-RIE) with CHF$_3$ gas. At last, a 3-$\mu$m layer of silica is deposited on the device via inductively coupled plasma chemical vapour deposition (ICP-CVD) as the cladding. Regarding this work, the ChG waveguides and the microresonators have a fixed height of 0.8 $\mu$m, and the width is flexibly tailored in the range 1.6-2.4 $\mu$m, which allows for dispersion engineering as well as engineered Raman lasing.

\section{Results and Discussions}

\subsection{ChG Microresonator Characterization}

A schematic of the GeSbS-based microring resonator is shown in Fig. \ref{fig:1}(a). The Raman gain spectrum of the GeSbS film was experimentally measured by Raman spectrometer, (see Fig. \ref{fig:1} (b)), which shows a broadband spectral profile compared with that of crystalline materials \cite{Hu:21}. In particular, the main Raman gain is centered at the frequency offset of 340 cm$^{-1}$ ($\sim$10.2 THz), with a linewidth of ca. 1.73 THz (estimated by a single-Lorentzian fit of the gain profile) \cite{MUSGRAVES20115032}. A complete analysis of all vibrational modes underlying the gain spectrum is presented in the supplementary information (SI). The normalized Raman peak gain is ca. 6.0 (extracted with $\int_0^\infty h_R(t) dt=1$, where $h_R(t)$ is the temporal Raman response function). The nonlinear coefficiency (\emph{n$_2$} ) of GeSbS is ca. $2.0\times10^{-18}$ m$^2$/W \cite{shang2021chip}. A typical SEM picture of the fabricated GeSbS microresonators is presented in Fig. \ref{fig:1}(e), together with the picture of the magnified coupling region. For the purpose of Raman lasing, the GeSbS waveguide is designed to support normal cavity dispersion. This will suppress nonlinear parametric oscillations usually evoked by the spontaneous Kerr nonlinearity in the anomalous dispersion regime \cite{suzuki2018broadband}. As such, engineering the waveguide geometry (i.e. changing the core width and height) will mainly contribute to tuning the FWM phase-matching condition among the pump wave and the Raman-induced Stokes / anti-Stokes waves. Consequently, there are mainly two types of the Raman lasing: single-mode and cascaded Raman lasers in the FWM phase-matching condition, and a relative broadband Raman-Kerr comb in the presence of phase mismatch, see Fig. \ref{fig:1}(c,d). In a microresonator with the FSR of $\sim$200 GHz, a typical resonance linewidth of $\sim$150 MHz was measured in the regime of the critical coupling, see Fig. \ref{fig:1}(i). This indicates a high intrinsic $Q$-factor ($Q_i$) >$2.2\times10^{6}$. In exploring the Raman lasing dynamic, the GeSbS microresonator is deliberately designed to support under coupled pump modes and over coupled Stokes modes, such that the conversion efficiency is effectively elevated at the transmitted port of the microresonator \cite{liu2018ultralow}.

\subsection{Low-threshold single-mode Raman laser}

We characterized the Raman lasing in the fabricated GeSbS microresonators. In the experiment, an amplified continuous wave (CW) tunable  laser is coupled to the resonator. The insertion loss of fiber-to-chip coupling is $\sim$2.5 dB/facet for the fundamental TE mode. At the output, the optical signal is monitored by an optical spectral analyzer (OSA), in which the generated Raman lasing is quantitatively noted (cf. SI for a detailed experimental setup).

At a low pumping power, the threshold for the Raman lasing is identified. In this condition, only single-mode Raman lasers are observed, see Fig. \ref{fig:2}(a). When the microresonator is pumped at 1550 nm with the power of 3.5 mW, a significant Stokes laser is monitored at 1637 nm, in the transmitted port of the resonator. The frequency offset is $\sim$10.3 THz, which is close to the central peak of the Raman gain. In addition, a weak anti-Stokes laser is also monitored at the blue side of the pump with a similar frequency offset. Moreover, the transmitted power of this single-mode Stokes wave is traced as a function of the pump power, in which a threshold power of $\sim$3.25 mW can be identified [Fig. \ref{fig:2}(b)]. Above the threshold, the Stokes wave is enhanced when the pump power is increased. The external slope efficiency is extracted to be 13.86$\%$. We can further deduce the Raman gain coefficient (g$_R$) of GeSbS material, which is ca. $7.37\times10^{-12}$ m/W (a value over 100 times higher than that in silica.)

\begin{figure}[ht!]
\centering\includegraphics[width=\linewidth]{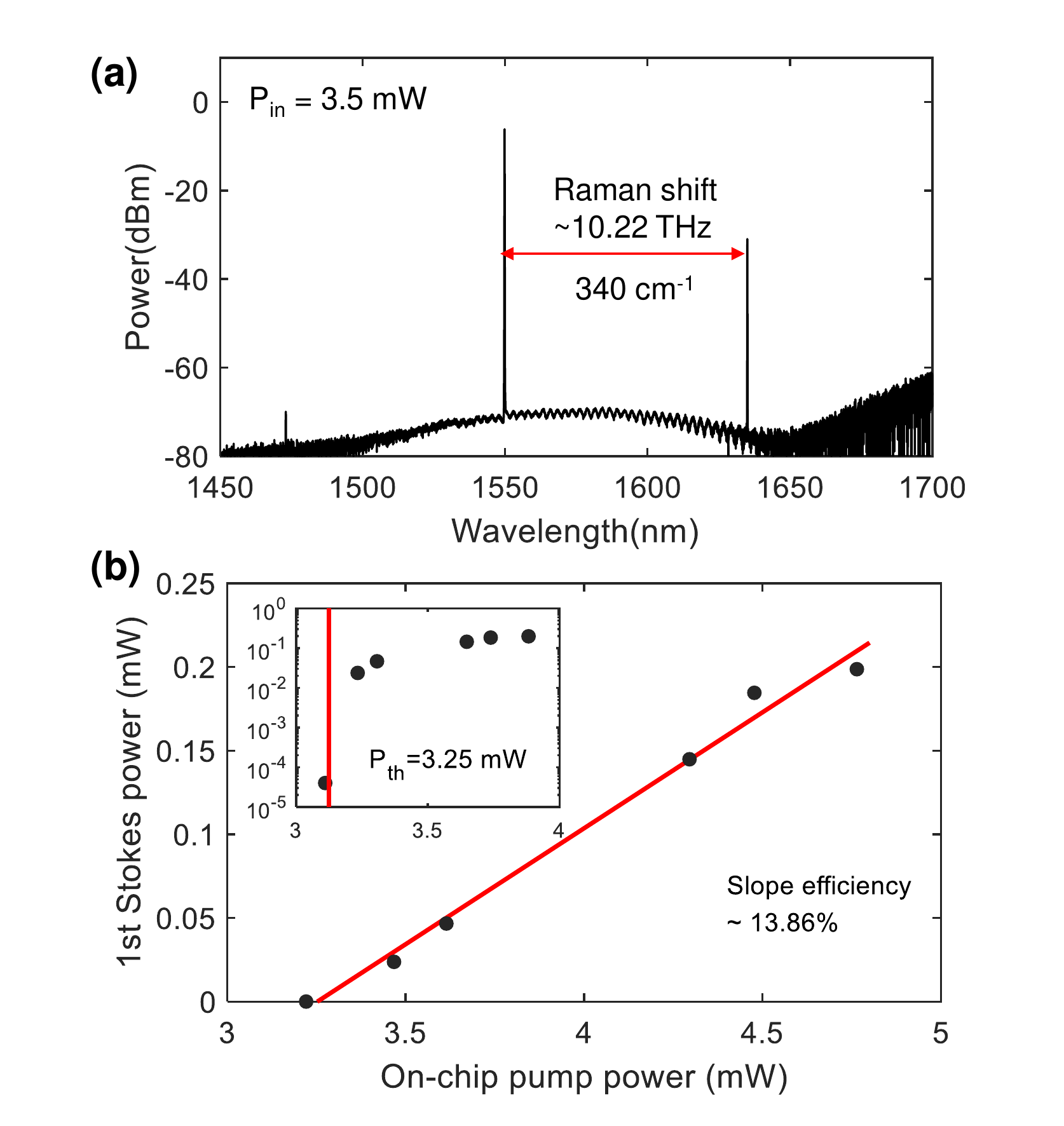}
\caption{\textbf{Raman threshold power.} (a) Measured optical spectra of the first Stokes wave at around 1637 nm, with the on-chip pump power estimated to be $\sim$3.5 mW. (b) The power trace of the first Stokes wave as a function of the pump power.The slope efficiency is $\sim$13.86$\%$ and the threshold power is ca. 3.25 mW.}
\label{fig:2}
\end{figure}

\begin{figure*}[ht!]
\centering\includegraphics[width=\linewidth]{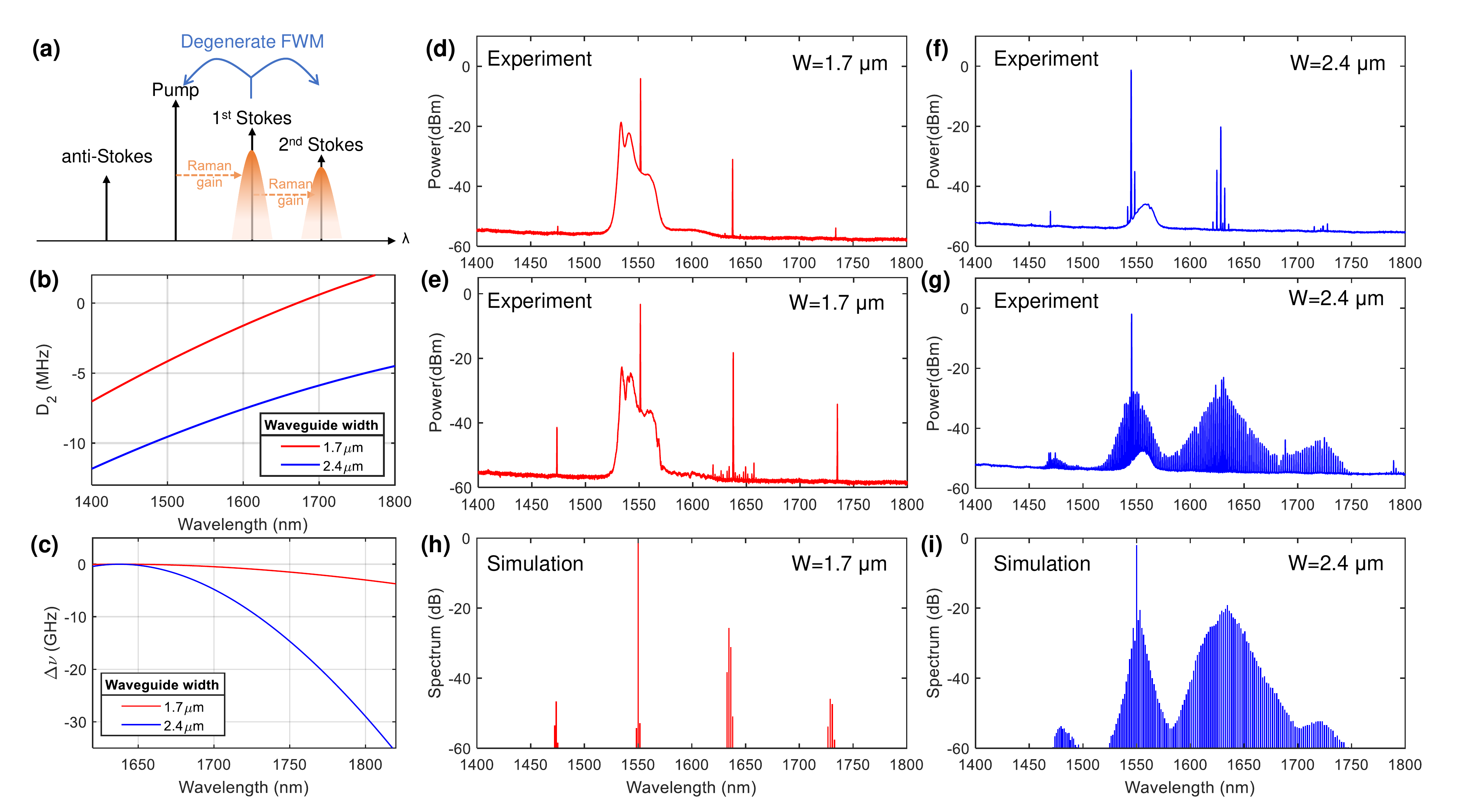}
\caption{\textbf{Engneered Raman lasing dynamics.} (a) Schematic of the degenerate FWM between the pump wave, the first and the second Stokes waves. (b) Calculated dispersion parameter $D_2$ of the microresonators, with different waveguide widths (1.7 $\mu$m and 2.4 $\mu$m). The $Q$ factors of the 1.7-$\mu$m microresonator is $Q_i=0.93\times10^6$, $Q_L=0.8\times10^6$, $Q_c=5.8\times10^6$. (c) Calculated phase mismatch for the degenerate FWM process ($\Delta\nu^1$). (d, e) The measured Raman spectra in the 1.7-$\mu$m microresonator, by narrowly tuning the laser frequency (i.e. changing the laser-cavity detuning). This operation is equivalent to the change on the cavity coupling power. The pump power is $\sim$35 mW (corresponding intracavity power intensity is 345.1 mW/cm$^2$). The first Stokes and second Stokes lasing with Raman shifts of $\sim$10.22 THz and $\sim$10.24 THz are generated almost simultaneously, implying similar threshold power. (f, g) The measured Raman spectra of the microring with 2.4 $\mu$m width by changing the detuning. The pump power is $\sim$20 mW (corresponding intracavity power intensity is 273.7 mW/cm$^2$). Numerical simulations in both dispersion conditions of the microresonators with the width of (h) 1.7 $\mu$m and (i) 2.4 $\mu$m.}
\label{fig:3}
\end{figure*}

\subsection{Engineered Raman Lasing at high pump power}

In the presence of a strong pump wave, the Raman lasing dynamic is enriched. Given that the nonlinear parametric oscillations are effectively suppressed (in the condition of overall normal cavity dispersion), dispersion engineering within a certain range would mainly alter the potential phase matching condition for the SRS as well as for the Raman lasing. In particular, with respect to the generation of the second Stokes wave, there are FWM processes along with the Raman effect. One is the degenerate FWM process involving the pump wave, the first and the second Stokes waves. In this case, the pump energy is converted to the first Stokes wave because of the accompanied SRS effect, followed by the conversion between the first and the second Stokes waves. This would lead to a strong second Stokes wave that may surpass the anti-Stokes wave in the power. The other is the non-degenerate FWM process involving the pump wave, the first and the second Stokes waves and the anti-Stokes wave. In microresonators, the phase mismatches ($\Delta\nu$) of these FWM processes are reflected by the difference of resonant frequencies of related modes \cite{Jean:20}, i.e.:
\begin{equation}
\Delta \nu^1 = 2\omega_{s,1} - \omega_0 - \omega_{s,2}
\label{eq:1}
\end{equation}
\begin{equation}
\Delta \nu^2 = \omega_0 + \omega_{s,1} - \omega_{as,1} - \omega_{s,2}
\label{eq:2}
\end{equation}
where $\omega_0$ indicates the resonant frequency of the pumped mode, $\omega_{s,1}$, $\omega_{s,2}$ are resonant frequencies of modes that support the first and the second Stokes waves, respectively, $\omega_{as,1}$ is the resonant frequency of the anti-Stokes mode. 

Clearly, the FWM phase-matching condition can be altered by tailoring the dispersion of the microresonator, which is implemented \emph{via} tuning the core width of the GeSbS waveguides. In our experiments, we designed mainly two types of GeSbS microresonators, with a difference in the cavity dispersion [see Fig. \ref{fig:3}(b)]. In detail, in microresonators with a waveguide width of 1.7 $\mu$m, the calculated cavity dispersion indicates that both of the phase mismatch values $\Delta\nu^1$ and $\Delta\nu^2$ are close to zero, while with a width of 2.4 $\mu$m, the phase mismatch is large [see Fig. \ref{fig:3}(c) and Fig. S-3]. 

As a result, distinguished Raman lasing dynamics are observed. In the case of small phase mismatch, the FWM processes will be dominant, leading to single-mode and cascaded Raman lasing when the pump power is increased [Fig. \ref{fig:3}(d, e)]. In contrast, in the case of large phase mismatch, a remarkable broadband comb spectrum is observed mainly around the pump wave and the first Stokes wave [see Fig. \ref{fig:3}(g, h)]. This is understood that when the energy conversion to the second Stokes wave is suppressed as being largely phase mismatched, the accumulated phase in the first Stokes wave would surpass the FSR of the resonator and start to convert energy to the neighboured resonant modes. As a consequence, the spontaneous Kerr nonlinear process is triggered, which leads to the generation of “sub-Kerr-combs” around the first Stokes wave and around the pump wave. The FWM phase-matching conditions at different pump wavelengths are also estimated, cf. the SI, in which the phase mismatch value ($\Delta\nu^1$) is slightly decreased with an increase in the pump wavelength. As such, a smooth transition from the single-mode cascaded Raman lasing to the Raman-Kerr comb state is observed, cf. the Raman lasing spectra presented in Fig. S-4.

Indeed, such a Raman-Kerr comb has been previously reported in whispering gallery mode crystalline resonators, which was also attributed to the cavity dispersion effect \cite{chembo2015spatiotemporal}. However, due to the fact that performing dispersion engineering in such resonators is difficult, the pump frequency has to be carefully selected to meet the required dispersion condition. In this context, our work may contribute to providing a flexible photonic platform that allows for engineered Raman lasing performance.

We also performed numerical simulations in both dispersion conditions underlying the two microresonator structures (i.e. the waveguide width of 1.7 $\mu$m and 2.4 $\mu$m). The simulation is based on the \emph{Lugiato-Lefever} equation with the complete form of the Raman response included \cite{chembo2015spatiotemporal}. As a result, both the cascaded Raman lasing and the Raman-Kerr comb are observed by simulations, which show good agreement with experiment results. In addition, simulations also indicate that the cascaded Raman lasing represents a stable state in the cavity, while the Raman-Kerr comb is a non-stable state.

\subsection{Tunability of Raman Laser}

We next demonstrated the tunability of the Raman Lasers in the FWM phase-matching condition, which is by means of tuning the pump wave or by tuning the operating temperature on the photonic chip. As a result, by consecutively couple the pump wave into the resonant modes of the microresonator, both the first and the second Stokes waves are shifted accordingly, see Fig. \ref{fig:4}(a, b). The tested tuning range for the first Stokes wave is 1615-1658 nm and that for the second Stokes wave is 1720-1755 nm, while the pump wavelength is tuned within the telecommunication $C$-band (1540-1565 nm).

Moreover, the Raman laser can also be continuously tuned by tuning the temperature. It is observed that with an increase in the temperature, the first Stokes wave is red shifted, see Fig. \ref{fig:4}(c,d).  By tuning the temperature of $\sim$40 ${}^{\circ}\mathrm{C}$, the first stokes wave is continuously swept over one FSR (ca. 220 GHz) of the microresonator.

Therefore, with a combination of both the pump laser and the temperature tuning schemes, single-mode Raman lasing at an arbitrary wavelength in the range 1615-1755 nm is accessible. In principle, such a tunability on the Raman lasing is unlimited since the Stokes wave is only coupled to the pump wave with a fixed frequency offset. Yet, at the moment, we are limited by the operational bandwidth of our laser sources as well as the optical fiber amplifiers. On another aspect, the FWM phase-matching condition should always be carefully controlled as it may drift over the wavelength and becomes detrimental to the laser tunability.

\begin{figure}[ht!]
\centering\includegraphics[width=\linewidth]{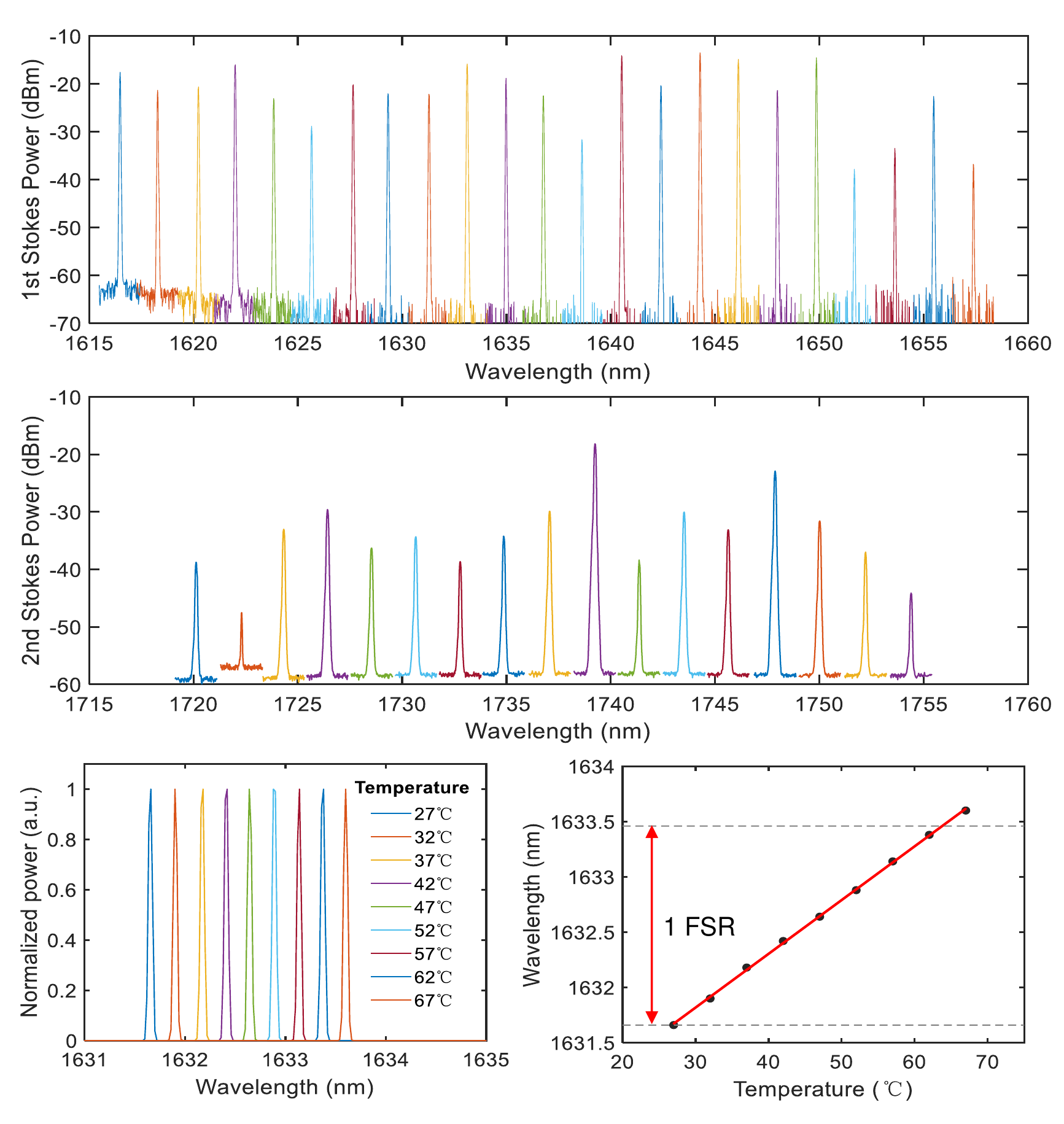}
\caption{\textbf{Tunability of the Raman laser.} Discrete tuning of (a) first Stokes and (b) second Stokes pumped at $C$-band boosted by an EDFA. (c) The pump frequency is tuned within a thermally red-shifted resonance. The output power is normalized to the peak emission at each pump wavelength. (d) Individually measured first Stokes wavelengths of the microring resonator corresponding to the different temperatures. Continuous tuning of the Stokes wavelength over $\sim$220 GHz (more than 1-FSR).}
\label{fig:4}
\end{figure}

\section{Conclusion}

In summary, we have presented an engineered Raman lasing in an integrated GeSbS microring resonator with a high $Q$-factor (above 10$^6$). We achieved the single-mode and the cascaded Raman lasing, together with a broadband Raman-Kerr comb, by tailoring the dispersion of the microresonator. We also provided an insightful understanding of the effect of dispersion on the interaction of the Kerr-Raman effect. Moreover, single-mode Raman lasing at an arbitrary wavelength in the range 1615-1755 nm is accessible by combining the pump laser tuning and the temperature tuning schemes. By comparing Raman laser characteristics based on different material platforms [Table S-1], our ChG-based Raman microresonator displays a relatively competitive advantages in the low pumping threshold (ca.3.25 mW) and high slope efficiency (ca. 13.6$\%$). Further improvement is expected by improving the intrinsic Q-factor and optimizing the coupling design. Our results pave the way for the generation of varietal on-chip Raman lasers and Kerr frequency combs covering the NIR to MIR range based on the ChG photonic platform.

\begin{backmatter}
\bmsection{Funding}National Key R$\&$D Program of China under Grant (2019YFA0706301), National Science Foundation of China (NSFC) (U2001601, 61975242, 61525502, 61435006, 61490715), the Science and Technology Planning Project of Guangdong Province (2019A1515010774), the Science Foundation of Guangzhou City (202002030103).

\smallskip


\bmsection{Disclosures} The authors declare no conflicts of interest.
\\
\\
\noindent See Supplement 1 for supporting content. 

\end{backmatter}




\ifthenelse{\equal{\journalref}{aop}}{%
\section*{Author Biographies}
\begingroup
\setlength\intextsep{0pt}
\begin{minipage}[t][6.3cm][t]{1.0\textwidth} 
  \begin{wrapfigure}{L}{0.25\textwidth}
    \includegraphics[width=0.25\textwidth]{john_smith.eps}
  \end{wrapfigure}
  \noindent
  {\bfseries John Smith} received his BSc (Mathematics) in 2000 from The University of Maryland. His research interests include lasers and optics.
\end{minipage}
\begin{minipage}{1.0\textwidth}
  \begin{wrapfigure}{L}{0.25\textwidth}
    \includegraphics[width=0.25\textwidth]{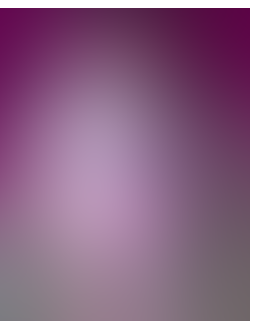}
  \end{wrapfigure}
  \noindent
  {\bfseries Alice Smith} also received her BSc (Mathematics) in 2000 from The University of Maryland. Her research interests also include lasers and optics.
\end{minipage}
\endgroup
}{}

\appendix
\clearpage
\onecolumn
\section{Engineered Raman Lasing in Photonic Integrated Chalcogenide Microresonators: supplemental document}

This document provides supplementary information to “Engineered Raman Lasing in Photonic Integrated Chalcogenide Microresonators,” including GeSbS Raman spectroscopy, experimental setup, detailed calculation of frequency mismatch, numerical simulations and material platforms comparison.

\subsection{Raman spectroscopy measurement}
The Raman spectrum of the GeSbS film deposited on a pure quartz substrate was measured at room temperature using the Raman Spectrometer (Renishaw inVia Reflex). The pump wavelength of the laser is 785 nm, see Figure. \ref{fig:1}. Lorentz fitting of the Raman spectra was carried out to distinguish the vibrational frequencies of the Raman-active phonon. The possible frequencies of molecular bond vibration modes of GeSbS films are listed in figure \cite{sakaguchi2019structural, musgraves2011comparison}, and the corresponding peaks are fitted by dotted lines in Fig.\ref{fig:1}.The main peak gain is located at around 330 cm$^{-1}$ (9.9 THz), corresponding to the vibrational frequencies of the GeS$_{4/2}$ bond\cite{sakaguchi2009structural}.

\begin{figure}[ht!]
\centering\includegraphics[scale=0.5]{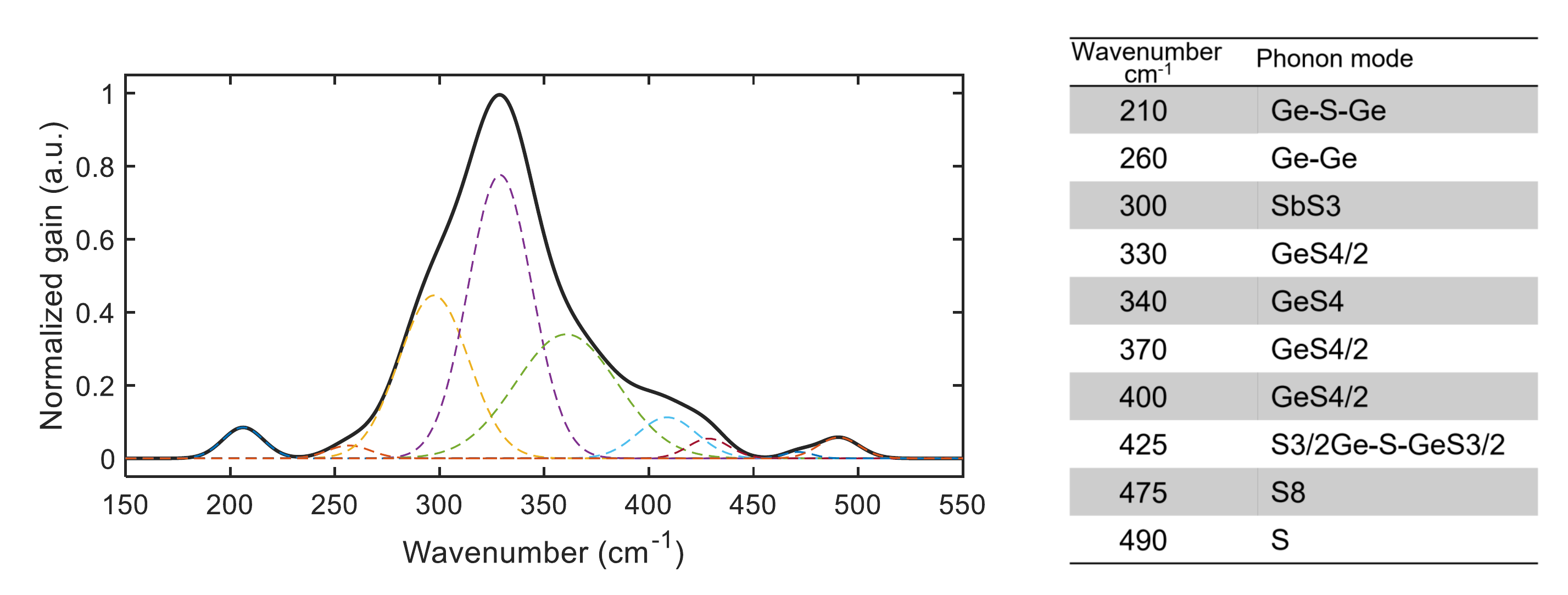}
\caption{Measured Raman Spectra of the GeSbS film and corresponding phonon modes.}
\label{fig:1}
\end{figure}

\subsection{Experimental setup for Raman lasers measurement}
The experimental setup for measuring the Raman lasers in GeSbS microrings is schematically illustrated in Fig. S2, similar to Ref \cite{suzuki2018broadband}. Briefly, a tunable continuous wave (CW) laser (Toptica CTL 1550) was used as the pump laser, which was amplified by an erbium-doped fiber amplifier (EDFA, Amonics AEDFA-33-B-FA). The polarization of pump light was aligned with fiber polarization controllers (FPCs) to TE polarization of the waveguide. The output power was monitored using Oscilloscope to realize the effective detuning between the pump laser and resonance. The results of Raman lasers spectra were recorded by two optical signal analyzers (OSA, Yokogawa AQ6370D with available wavelength range from 0.7$\mu$m to 1.7$\mu$m and a wavelength resolution of 20pm, and AQ6375B with available wavelength range from 1.2$\mu$m to 2.4$\mu$m a wavelength resolution of 50 pm) ). The temperature controller (TEC) is applied to keep thermal stability of the on-chip microrings during the Raman laser measurement.

\begin{figure}[ht!]
\centering\includegraphics[scale=0.6]{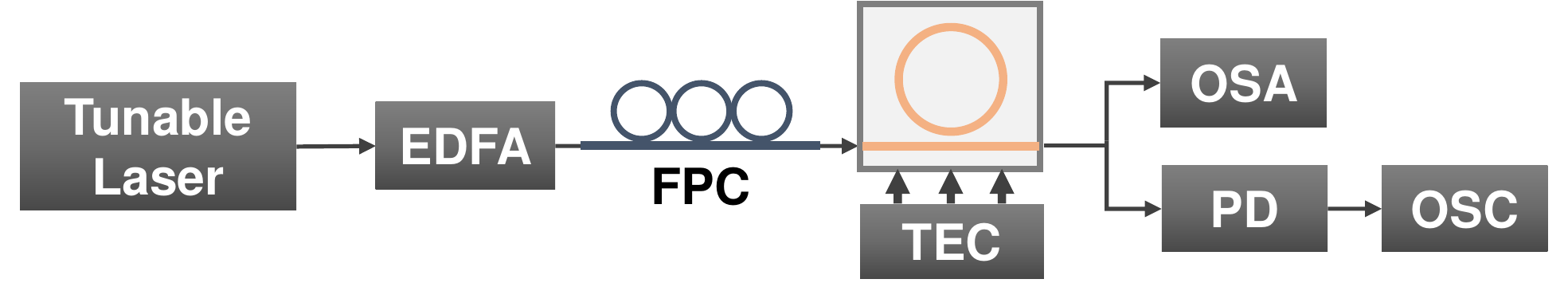}
\caption{Schematically illustrating the experimental setup for Raman lasers generation.}
\label{fig:2}
\end{figure}

\begin{figure}[ht!]
\centering\includegraphics[scale=0.6]{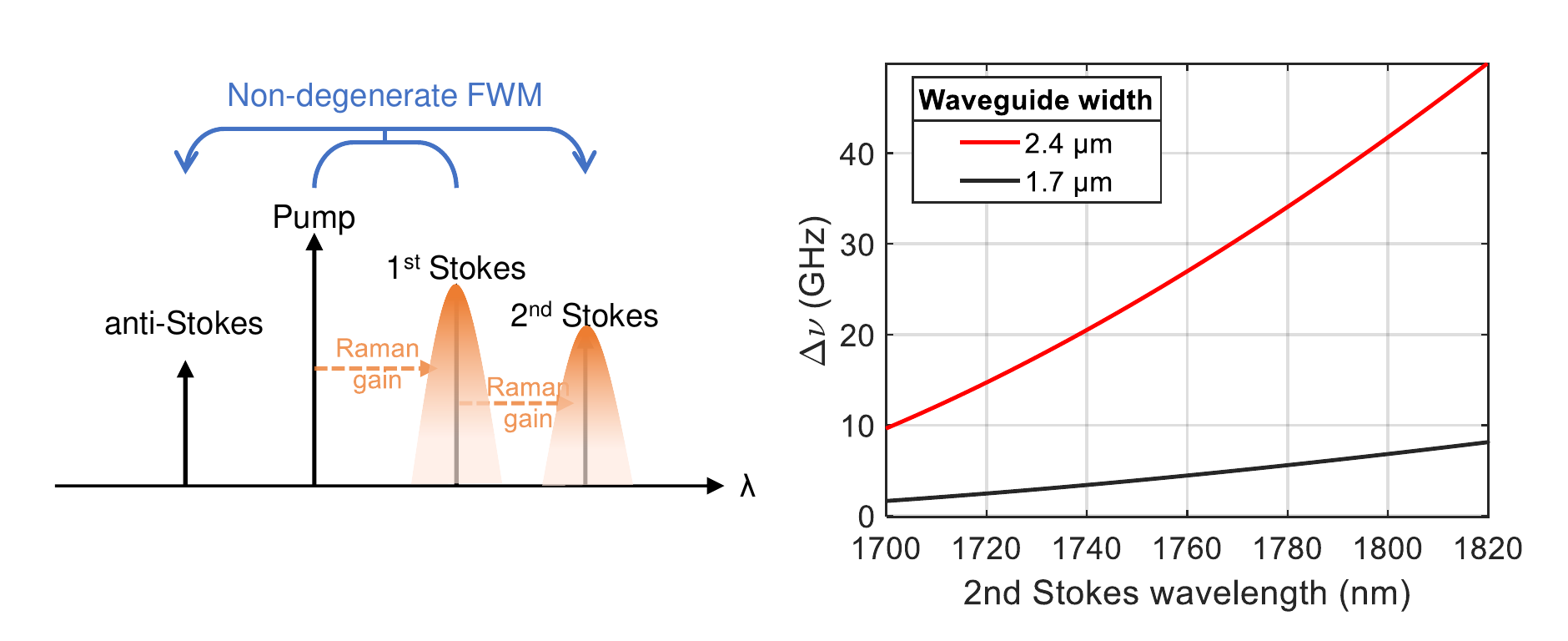}
\caption{(a)A schematic illustrating the non-degenerate FWM between the pump,1st Stokes, 2nd
Stokes and anti-Stokes modes.(b)Calculated the frequency mismatch value of non-degenerate FWM for the two microrings with the width of 1.7 $\mu$m and 2.4 $\mu$m.}
\label{fig:3}
\end{figure}

\subsection{Frequency matching for FWM process}
Beside the frequency mismatching of degenerated FWM discussed in the main text, here, we calculated the frequency mismatching values of the non-degenerate FWM process involving the pump wave, the first and the second Stokes waves and the anti-Stokes wave, see Fig. \ref{fig:3}. In this case, the calculated cavity dispersion indicates that the phase mismatch values ($\Delta \nu$) in microresonators with a waveguide width of 1.7 $\mu$m is also close to zero, while the phase mismatch is large in the microring with a width of 2.4 $\mu$m. 

Moreover, the different phase mismatch values lead to distinguished dynamics of the Raman lasing when the microresonators are pumped at different waves. The phase mismatch values of both microrings with the pump waves ranging from 1530 nm to 1560 nm are shown in Fig.\ref{fig:4}(a, b). Markedly, in the microring with a width of 1.7 $\mu$m, the $\Delta\nu$ is increasing at shorter pump waves, leading to the suppressed cascaded Raman lasering. As a result, the spontaneous Kerr nonlinear process is triggered, which leads to the generation of “sub-Kerr-combs” around the first and the second Stokes waves when pumped at 1541.5 nm [Fig. \ref{fig:4}(b)], while pumped at 1560.2 nm, the typical cascaded Raman laser could be observed [Fig. \ref{fig:4}(c)]. For the microring with a width of 2.4 $\mu$m, the frequency mismatch values are all large for different pump waves, see Fig. \ref{fig:4}(d), leading to the similar Raman combs when pumped at both 1540.4 nm and 1563.2 nm [Fig. \ref{fig:4}(e, f)]. Such phenomena further illustrate the frequency mismatching of the FWM plays a significant role in the SRS as well as the Raman lasing in our GeSbS microresonators.

\begin{figure}[ht!]
\centering\includegraphics[width=\linewidth]{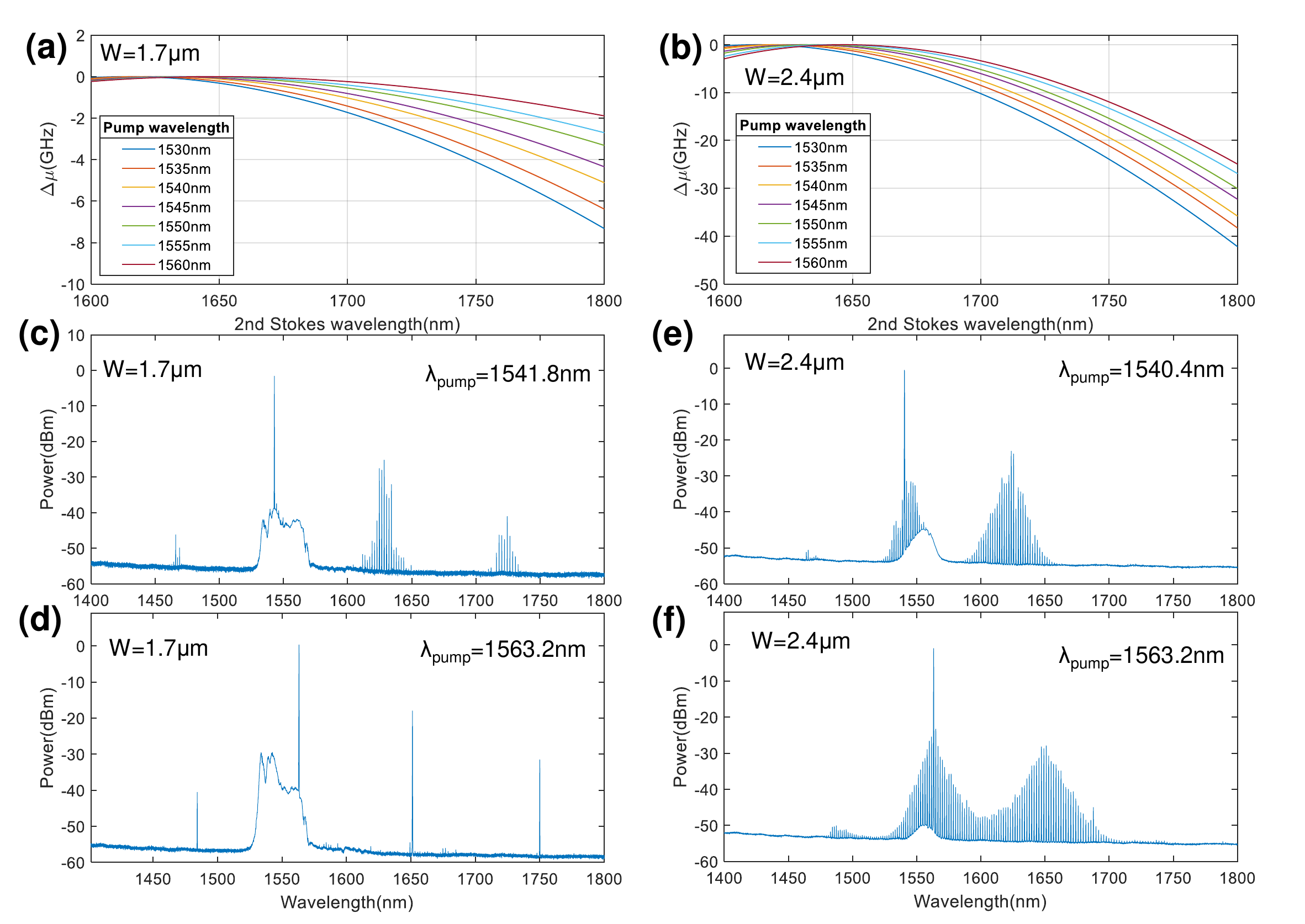}
\caption{Calculated frequency mismatch values of the degenerate FWM process between the pump wave, the first and the second Stokes waves for the microring with a width of (a) 1.7 $\mu$m and (b)2.4 $\mu$m, respectively, at various pump wavelengths from 1535 nm to 1560 nm. The corresponding first Stokes is spanning from around 1613 nm to 1650 nm for both microrings. Measured Raman spectra for the microring with 1.7 $\mu$m width when pumped at (c) 1541.8 nm, (d) 1563.2 nm with the same pump power of 35 mW, respectively. Measured Raman spectra for the microring with 2.4 $\mu$m width when pumped at (e) 1540.4 nm, (f) 1563.2 nm with the same pump power of 20 mW, respectively.}
\label{fig:4}
\end{figure}

\subsection{Numerical simulations}
The following section introduces the modelling of the Raman-Kerr comb in GeSbS microresonators. The master equation reads:

\begin{equation}
\begin{aligned}
\frac{\partial}{\partial t} A_{\mu}=&\left(-\frac{\kappa}{2}-i \delta_{\omega}-i D_{\text {int } \mu}\right) A_{\mu}+\sqrt{\kappa_{e x}} s_{i n} \cdot \delta_{\mu} \\
&-i g_{\mu} F\left[\left(1-f_{R}\right)|A(t, \theta)|^{2} A(t, \theta)+ \right.\\
&\phantom{=\;\;}\left.f_{R} \cdot F^{-1}\left[h_{R}(\mu) \cdot F\left[A^{2}(t, \theta)\right]_{\mu}\right] A(t, \theta)\right]
\label{eq:1}
\end{aligned}
\end{equation}

where A$_{\mu}$ is the complex amplitude with respect to the resonant mode $\mu$, which in the Fourier domain reads A(t, $\theta$) ($\theta$ stands for a fast axis corresponding to the intracavity phase domain $\theta \in[-\pi, \pi)$ and t is the slow-time frame), $\kappa$ indicates the resonance linewidth that consists of both the intrinsic loss rate $\kappa_{0}$ and the coupling loss rate $\kappa_{ex}$, $\delta_{\omega}=\omega_{0}-\omega_{p}$ is the laser- cavity detuning ($\omega_{0}$ is the angular frequency of the pumped mode and $\omega_{p}$ is that of the pump wave), D$_{int,\mu}$ is the integrated dispersion profile, g$_{\mu}$ is the coefficient of the nonlinear-induced frequency shift, h$_{R}$($\mu$) indicates the normalized Raman spectrum in $\mu$-domain, f$_{R}$ is the Raman fraction, s$_{in}$ is the external pump field and $\delta_{\omega}$ is the Kronecker delta function.

\begin{figure}[ht!]
\centering\includegraphics[width=\linewidth]{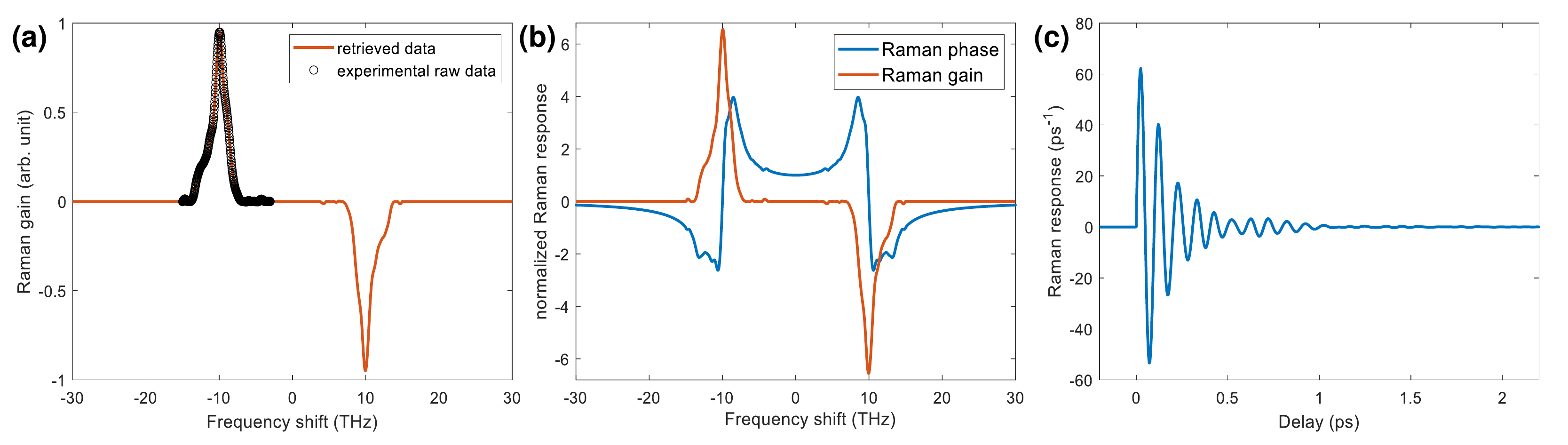}
\caption{(a) Measured and retrieved Raman gain spectrum; (b) Normalized complete Raman spectrum, consisting of both the real part as the Raman phase spectrum and the imaginary part as the Raman gain spectrum; (c) Normalized Raman temporal response function.}
\label{fig:5}
\end{figure}

Note that without the Raman effects, this equation (also called the coupled mode equation\cite{chembo2010modal}) in the $\theta$ -domain is known as the Lugiato-Lefever equation\cite{lugiato2018lugiato}. The Raman spectrum is obtained by the experimental measure of the Raman gain spectrum $\left(\operatorname{Im}\left[h_{R}(\omega)\right]\right)$ and by the
normalization on the corresponding temporal response, i.e. $\int_{0}^{\infty} h_{R}(t) d t=1$, see Fig.\ref{fig:5}. In
particular, the Raman phase spectrum $\left(\operatorname{Re}\left[h_{B}(\omega)\right]\right)$ can be calculated from the gain spectrum by using the Cauchy principal value, i.e.:
\begin{equation}
    \operatorname{Re}\left[h_{R}(\omega)\right]=\frac{1}{\pi} p . v \cdot \int_{-\infty}^{+\infty} \frac{\operatorname{Im}\left[h_{R}\left(\omega^{\prime}\right)\right]}{\omega-\omega^{\prime}} d \omega^{\prime}
    \label{eq:2}
\end{equation}

\begin{figure}[ht!]
\centering\includegraphics[width=\linewidth]{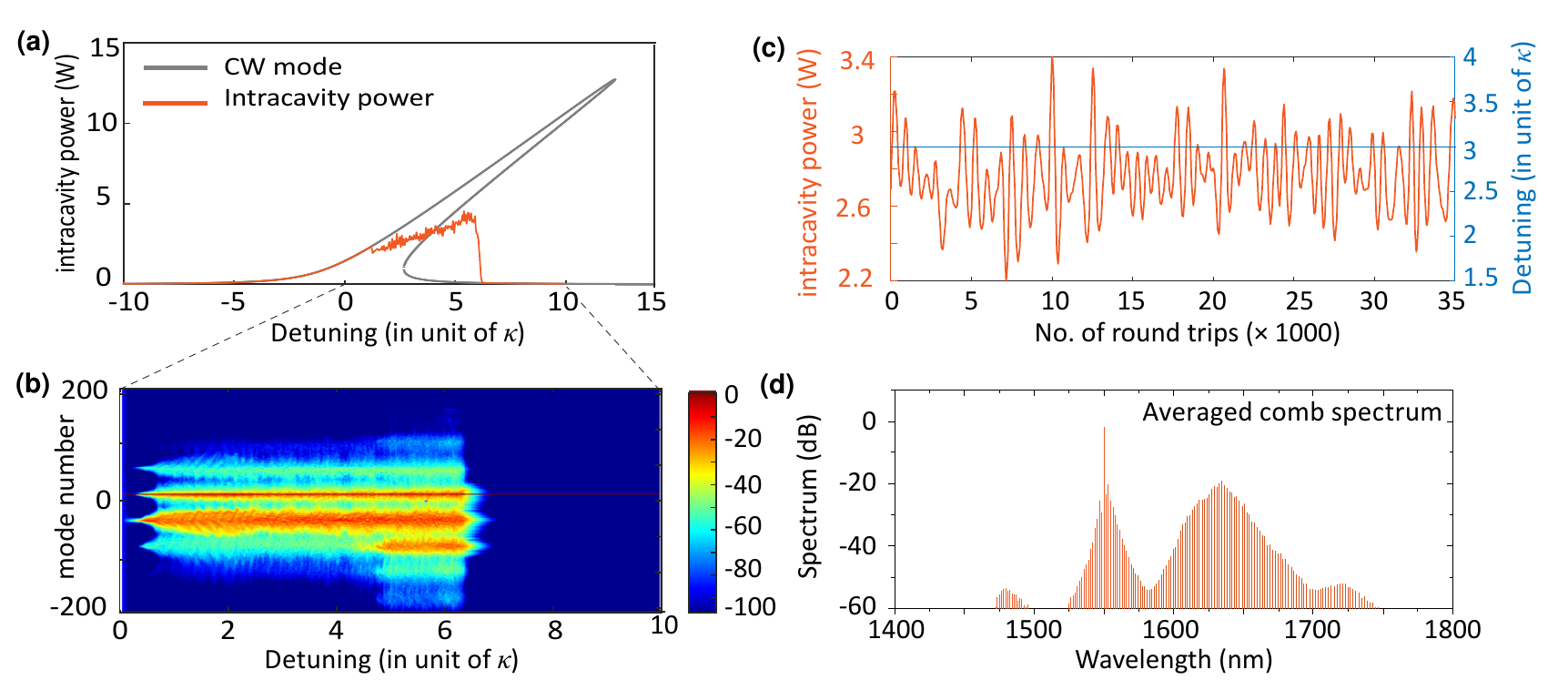}
\caption{(a) Modelled intracavity power trace upon a laser scan process, which is compared with a tilted resonance profile in the CW mode. (b) The evolution of the intracavity field spectrum over the laser detuning, which indicates broadband Raman lasing at Stokes / anti-Stokes wavelengths. (c) The power trace of the self-evolving intracavity field, with a fixed detuning value of $3 \kappa$, which implies an unstable filed pattern. (d) Transmitted spectrum showing a broadband Raman-Kerr comb, which is obtained by averaging a number $(35,000)$ of intracavity field patterns. In simulations, the second and the third order dispersion components are $D_{2}=-2 \pi \times 5$ MHz, $D_{3}=-2 \pi \times 22$ KHz, respectively, which is effective for the microresonator with a waveguide width of $2.4 \mu \mathrm{m}$. The pump power is  $30 \mathrm{~mW}$.}
\label{fig:6}
\end{figure}

In simulations, the laser scan process is modelled. The laser-cavity detuning is swept over the pumped resonance, and the averaged intracavity power is traced and is compared with a standard tilted resonance in the absence of nonlinear parametric processes (namely in the CW mode). Upon this process, the intracavity power trace starts to deviate from the CW mode when the detuning is slightly above zero (in the red-detuned regime), see Fig. \ref{fig:6}(a), indicating the generation of light fields at new frequencies. Indeed, in the spectrogram of the intracavity field, broadband Raman lasing at Stokes / anti-Stokes frequencies are observed, see Fig. \ref{fig:6}(b). By fixing the laser detuning, the self-evolution of the intracavity field is also modelled, whose power trace indicates that the broadband Raman lasing is an unstable state, see Fig. \ref{fig:6}(c). The transmitted field spectrum is then obtained by averaging a number of the intracavity field patterns, see Fig. \ref{fig:6}(d), which correctly reflects the nature of the slow spectral measurement in experiments by using an optical spectral analyzer. Meanwhile, the dynamics of cascaded Raman lasing can be simulated using this model with a different dispersion parameter of $D_{2}=-2 \pi \times 1.5$ MHz, $D_{3}=-2 \pi \times 26$ KHz.

\subsection{comparison of various material platforms for Raman laser generation}
We compared the performance of Raman lasers generated in both non-integrated and integrated platforms, see \ref{table1}. Despite the Raman laser in silica microtoroid shows both ultralow threshold and high conversion efficiency, the non-integrated feature hinders its practical application with high stability and compactness. In the integrated platforms, the threshold power of our GeSbS microring is comparable with the silicon racetrack resonator. The conversion efficiency is the same order of magnitude compared with silicon racetrack resonator, AlN microring and LNoI microresonator. We note that the performance of our GeSbS microring based Raman laser can be further improved by optimizing the quality factor, coupling coefficient and mode volume, which is of vital importance to the realization of an ultra-efficient Raman laser and Raman comb based on ChG microresonators.

\begin{table}[htbp]
\centering
\caption{\bf Performances of various material platforms for Raman laser generation}
\begin{tabular}{p{3cm}<{\centering}p{1cm}<{\centering}p{1cm}<{\centering}p{2cm}<{\centering}p{2cm}<{\centering}c}
\hline
Platform & $Q_{int}(10^{6})$ & {Threshold power(mW)} & Slope efficiency & Coupling system & Ref\\
\hline
Silicon photonic crystal & 4&0.001 & 8\% & Non-integrated & \cite{takahashi2013micrometre}\\
Silica microtoroid &100&0.074 & 45\% & Non-integrated & \cite{kippenberg2004ultralow}\\
Si-O-Si monolayer grafted silica microtoroid & >10 & 0.2 & 40\% & Non-integrated & \cite{shen2020raman}\\
PDMS-coated silica microsphere & >10 & 1.3 & 0.1\% & Non-integrated & \cite{li2013low}\\
As$_{2}$S$_{3}$ microsphere &>70& 0.37 & 0.013\% & Non-integrated & \cite{andrianov2021tunable}\\
Silicon racetrack resonator & 2 &1.3 & 1.1\% & Integrated & \cite{zhangraman}\\
Silicon racetrack resonator & 1.38 & 20 & 28\% & Integrated & \cite{zhangraman}\\
AlN microring & 2.3& 8 & 3.6\% & Integrated & \cite{liu2017integrated}\\
AlN microring & 1.5 &34 & 15\% & Integrated & \cite{liu2017integrated}\\
LNoI microresonator& 1.5 &20 & 46\% & Integrated & \cite{yu2020raman}\\
Diamond microring&0.44 &85 & 0.43\% & Integrated & \cite{latawiec2015chip}\\
\bf GeSbS microring &\bf 2.2 &\bf 3.5 & \bf 13.86\% & \bf Integrated & \bf This work\\

\hline
\end{tabular}
  \label{table1}
  
\end{table}

\bibliographystyle{osajnl}
\bibliography{sample_all}

\end{document}